\documentclass[pra,twocolumn,showpacs,amsmath,amssymb,superscriptaddress]{revtex4-1}

\usepackage[utf8]{inputenc}
\usepackage{eucal}
\usepackage{dsfont}
\usepackage{braket}\usepackage{amsthm}\usepackage{bbm}\newcommand{\ens}[0]{\ensuremath}
\newcommand{\iE}[0]{\ens{\mathrm{i}}}
\newcommand{\Eins}[0]{\ens{\mathbbm{1}}}
\usepackage[varg]{txfonts}
\usepackage{graphicx}

\begin{document}
\title{Spontaneous Photon Emission in Cavities}

\author{G. Alber}

\affiliation{Institut für Angewandte Physik, Technische Universität Darmstadt,D-64289,
Germany}
\author{N. Trautmann}

\affiliation{Institut für Angewandte Physik, Technische Universität Darmstadt,D-64289,
Germany}
\date{December 4, 2014}
\begin{abstract}
 We investigate spontaneous photon emission processes of two-level atoms in parabolic and ellipsoidal cavities thereby taking into account the full multimode scenario.
In particular, we calculate the excitation probabilities of the atoms and the energy density of the resulting few-photon electromagnetic radiation field by using semiclassical methods for the description of the multimode scenario. Based on this
approach
photon path representations are developed for relevant transition probability amplitudes which are valid in
the optical frequency
regime where the dipole and the rotating-wave approximations apply.
Comparisons with numerical results demonstrate the quality of these semiclassical results even in cases 
in which the wave length of a spontaneously emitted photon becomes comparable or even
larger than characteristic length scales of the cavity. This is the dynamical regime in which
diffraction effects become important so that geometric optical considerations are
typically not applicable.
\end{abstract}
\maketitle
\section{Introduction}
The investigation of resonant light-matter interaction has received considerable attention during the last decades and remarkable experimental progresses in this field has opened the door to whole new experimental opportunities \cite{Berman1994,Walther2006,Haroche2006}. These developments
are not only interesting from a fundamental point of view but also from practical points of view for possible
applications in the area of quantum
information processing, for example. In this latter context methods for achieving
an optimal transfer of quantum information 
between single photons (flying qubits)
and elementary material systems (stationary qubits) 
over large distances by using the resonant strong coupling, as required for the realization of a quantum repeater,
are of particular interest. 
Therefore, recently considerable efforts \cite{Goy1983,Meschede1985,McKeever2003} have been devoted to investigate the interaction of matter qubits with few
modes of the radiation field within the framework of Jaynes-Cummings-Paul models \cite{Schleich}. Recent developments stimulated extensions of this work to extreme
multimode scenarios or to scenarios involving structured continua of electromagnetic field modes
which are characteristic for half open cavities of a parabolic shape \cite{Maiwald2009,Maiwald2012}, for example.

In the following we review some of these recent research activities focusing on
the resonant matter-field interaction of two-level atoms located around the focal point of a parabolic shaped cavity
and generalize them to prolate ellipsoidal cavities which include parabolic cavities as limiting cases.
A particularly interesting elementary quantum electrodynamical situation arises in a prolate ellipsoidal cavity 
if two two-level atoms are trapped in the two foci of this cavity and exchange a single photon which is emitted spontaneously
by one of these two-level atoms.

\section{The quantum electrodynamical model}
In our model we investigate the dynamics of $n\in\left\{ 1,2\right\}$ identical two-level atoms with transition frequencies $\omega_{eg}$ and dipole matrix elements $\vec{d}_{a}=\bra{e}_{a}\hat{\mathbf{d}}_{a}\ket{g}_{a}$. These two-level atoms are
assumed to be situated in the focal points of a prolate elliptical cavity $\vec{x}_a$ and to interact
with the quantized radiation field inside this cavity.
Furthermore,  we assume that the sizes of the relevant electronic states two-level atoms are small compared to the wavelengths $\lambda_{\text{eg}} = 2\pi c/\omega_{eg}$ of the atomic transition 
so that the dipole approximation is applicable ($c$ is the speed of light in vacuum.).
Moreover, as we are considering atomic transitions in the optical frequency regime 
the relevant
coupling (Rabi-) frequencies are small in comparison with $\omega_{eg}$ so that
the rotating-wave approximation is applicable. Thus, the quantum electrodynamical interaction between
$n=2$ two-level atoms and the quantized electromagnetic field inside the cavity
can be described by the
Hamiltonian
 $\hat{H}=\hat{H}_{\text{atoms}}+\hat{H}_{\text{field}}+\hat{H}_{\text{i}}$
with 
$\hat{H}_{\text{field}}=\hbar\sum_{i}\omega_{i}\hat{a}_{i}^{\dagger}\hat{a}_{i}$, $\hat{H}_{\text{atoms}}=\hbar \omega_{eg}\sum\limits_{a=1}^{n}\ket{e}_{a}\bra{e}_{a}\;,$
$\hat{H}_{\text{i}}=-\sum\limits_{a=1 }^{n}\hat{\mathbf{E}}_{\perp}^{+}(\vec{x}_{a})\cdot\hat{\mathbf{d}}_{a}^{-}+\text{H.c.}$
and with the dipole operator  $\hat{\mathbf{d}}_{a}^{-}=\vec{d}_a^{*}\ket{g}_{a}\bra{e}_{a}$ of the two-level atom $a$ .
Thereby, the energies of the atomic ground states $\ket{g}_{a}$ 
have been set equal to zero.
In the Schr\"odinger picture the positive frequency part of the
(transversal) electromagnetic field operator is given by
$$\hat{\mathbf{E}}_{\perp}^{+}(\vec{r}):=\iE\sum_{i}\sqrt{\frac{\hbar\omega_{i}}{2\epsilon_{0}}}\vec{g}_{i}(\vec{r})\hat{a}_{i}^{\dagger}$$
and  $\vec{g}_{i}(\vec{r})$ denote the normalized transversal electric mode functions which fulfill the boundary conditions for an ideally conducting cavity.

\section{Mode functions of the cavity and their semiclassical approximation}
For arbitrary boundary conditions the calculation of the mode functions  
$\vec{g}_{i}(\vec{r})$ is generally difficult. However, in cases in which the wave lengths of relevant field modes are small in comparison with all 
other length scales of the problems this task may be facilitated by semiclassical methods.
In order to put the problem into perspective let us consider cases in which the dipole matrix elements $\vec{d}_{a}$ of the atomic transitions are oriented along the symmetry axis, i.e. the z-axis, of a parabolic or ellipsoidal cavity.
Thus, we can write $\vec{d}_{a}=D \vec{e}_{z}$. 
This symmetry can be exploited as it is possible to show that in these cases
the only field modes which couple to the two-level atoms in the dipole approximation
are of the form
\cite{Moon1961}
$$\vec{g}_{i}(\varphi,\xi,\eta)=\nabla\times H,\;H=\vec{e}_{\varphi} F_i(\xi) G_i(\eta).$$
Thereby, $\varphi, \; \xi \; \text{and} \; \eta$ are the symmetry adapted
parabolic coordinates in the case of a parabolic cavity
and the prolate ellipsoidal coordinates in the case of an ellipsoidal cavity. 
For a parabolic cavity the functions $ F_i(\xi)$ and $G_i(\eta)$ turn out to be Coulomb functions and for an ellipsoidal cavity they turn out to be prolate spheroidal wave functions.
In both cases we can use semiclassical methods to express the quantization conditions resulting from the boundary conditions of an ideally conducting wall, for example, semiclassically in a simple way by taking advantage of the
separability of the Helmholtz equation in these coordinates. This way one obtains
a single quantization function for one of the coordinates, say $\eta$, in the case of a parabolic cavity and two quantization
conditions in the case of an ellipsoidal cavity. Furthermore, these semiclassical methods are also well suited
for normalizing these mode functions appropriately.	

\section{Photon path representation of probability amplitudes}
We can now use these results to calculate the time evolution of the system. Our Hamiltonian has the property that the number of excitations in the system (so $\sum\limits_{a=1}^{n}\ket{e}_{a}\bra{e}_{a}+\sum\limits_{i}a_{i}^{\dagger}a_{i}$) is a conserved quantity.
If we assume that initially only one two-level atom is in an excited state and that the field is in the vacuum state we know that for all later times the number of excitations in the system equals unity. This leads to the following ansatz for the wave function
$$\ket{\psi(t)}=b^{1}(t)\ket{e,g}^A\ket{0}^P+b^{2}(t)\ket{g,e}^A\ket{0}^P$$
$$+\sum\limits _{i}f_{i}(t)\ket{g,g}^A\ket{1}_{i}^P$$
in the case of two two-level atoms in the cavity. 
(The superscripts $A$ and $P$ refer to atoms and photons, respectively.)
It is a straight forward task to generalize the following calculation to an arbitrary number of atoms. The Schrödinger equation leads to a coupled system of linear differential equations. In order to solve this system we apply the Laplace transform in order to obtain a system of linear algebraic equations.
Due to basic properties of our Hamiltonian it is straight forward to eliminate the Laplace transforms of the photonic excitations $\widetilde{f}_{i}(s)$ ($s$ denotes the parameter of the Laplace transform) in order to
obtain a linear system of equation for $\widetilde{b}^{1}(s)
$ and $\widetilde{b}^{2}(s)$. This system of equations is given by
\begin{small}
$$
\left(\begin{array}{c}
b^{1}(0)\\
b^{2}(0)
\end{array}\right)=
\left[(s+\iE\omega_{eg})\Eins+\left(\begin{array}{cc}
A^{1,1} & A^{1,2}\\
A^{2,1} & A^{2,2}
\end{array}\right)\right]\left(\begin{array}{c}
\widetilde{b^{1}}(s)\\
\widetilde{b^{2}}(s)
\end{array}\right)
$$
\end{small}
with 
$$A^{a,b}:=|D|^{2}\sum\limits _{j}\frac{\omega_{j}}{2\epsilon_{0}\hbar}\frac{\left(\vec{g}_{j}(\vec{x}_{a})\right)_{z}\left(\vec{g}_{j}(\vec{x}_{b})\right)_{z}^{*}}{s+\mathrm{i}\omega_{j}}\;~~a
,b\in\left\{ 1,2\right\}.$$
The functions $A^{a,b}$ encode all the properties of the cavity. It is a non trivial task to evaluate these functions because they are defined by an infinite sum over all modes $j$ which couple to the atomic dipoles. However, we can evaluate these functions 
with the help of the semiclassical methods developed previously. With these functions at hand we can solve the linear system of equations and apply the inverse Laplace transform in order to obtain expressions for the time evolution of the probability amplitudes of observing atoms $1$ and $2$ in the excited states, i.e. $b^{1}(t)$
and $b^{2}(t)$. However, evaluating the inverse Laplace transform is still a non trivial task. For this purpose it is convenient to solve the linear system of equations by applying the Neumann series in order to obtain an infinite sum of simple expressions for
$\widetilde{b^{1}}(s)$ and $\widetilde{b^{2}}(s)$ and then apply the inverse Laplace transform to each of these terms. The Neumann series is given by
\begin{small}
$$\left(\begin{array}{c}
\widetilde{b^{1}}(s)\\
\widetilde{b^{2}}(s)
\end{array}\right)=\sum_{n=0}^{\infty}\frac{\left[-\left(\begin{array}{cc}
T_1^{1,1} & T_1^{1,2}\\
T_1^{2,1} & T_1^{2,2}
\end{array}\right)\right]^{n}}{(s+\frac{\Gamma_{\text{free}}}{2}+\iE \omega_{eg})^{n+1}}\cdot\left(\begin{array}{c}
b^{1}(0)\\
b^{2}(0)
\end{array}\right).$$\end{small}Thereby, we have defined $T_1^{a,b}:=A^{a,b}-\delta_{a,b}
\Gamma_{\text{free}}/{2}$ and $\Gamma_{\text{free}}$ denotes the spontaneous decay rate of each two-level system originating from spontaneous emission of a photon in free space, i.e. in the absence of any boundary conditions for the radiation field.
Of course, at first sight it may appear as a major complication to apply the Neumann series in order to invert a 2 by 2 matrix. However, this expansion leads
to a semiclassical photon path representation of the relevant probability amplitudes \cite{Milonni1974,Alber2013} so that we are able to interpret each term of this expansion as a contribution of a photon path. Thereby, each photon path involves a sequence of spontaneous emission and absorption processes.
In this expansion the functions $A^{a,b}$ encode the possible paths a photon can take to reach atom $b$ starting from atom $a$. In case of a parabolic cavity we can perform the same
calculations but instead of a two by two matrix of the functions $A^{a,b}$ we only have to consider the single term $A^{1,1}$.

\section{Modification of the spontaneous decay rate by boundary conditions}
\begin{figure}
	\centering
		 \includegraphics[width=5.5cm]{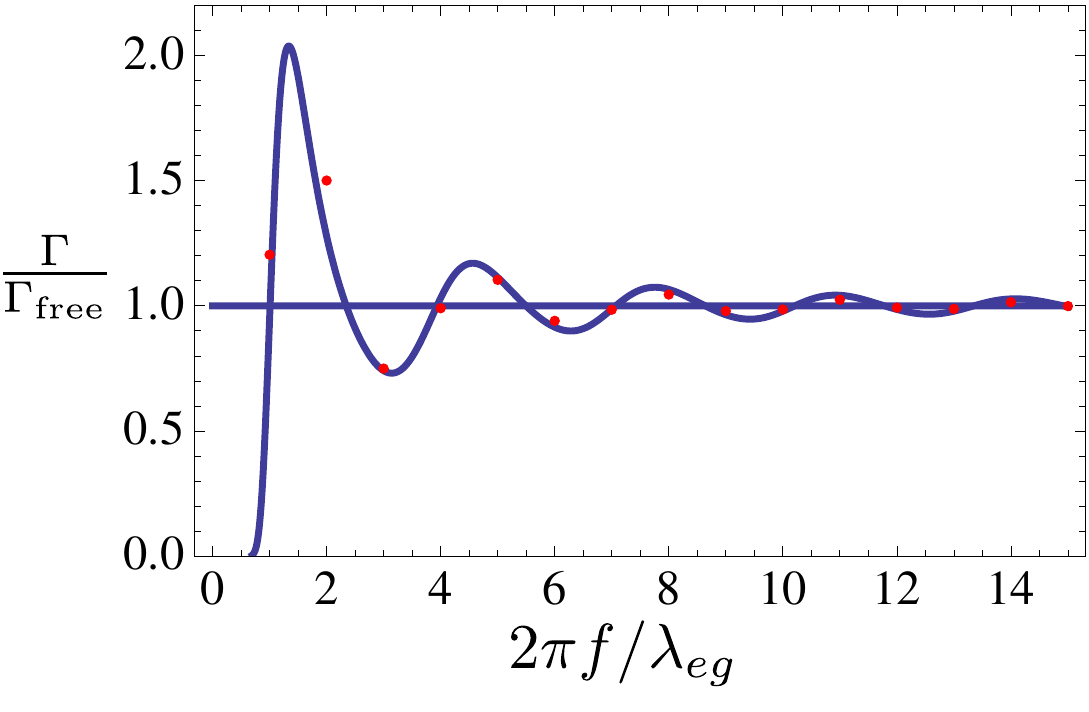}\\
		 \includegraphics[width=5.5cm]{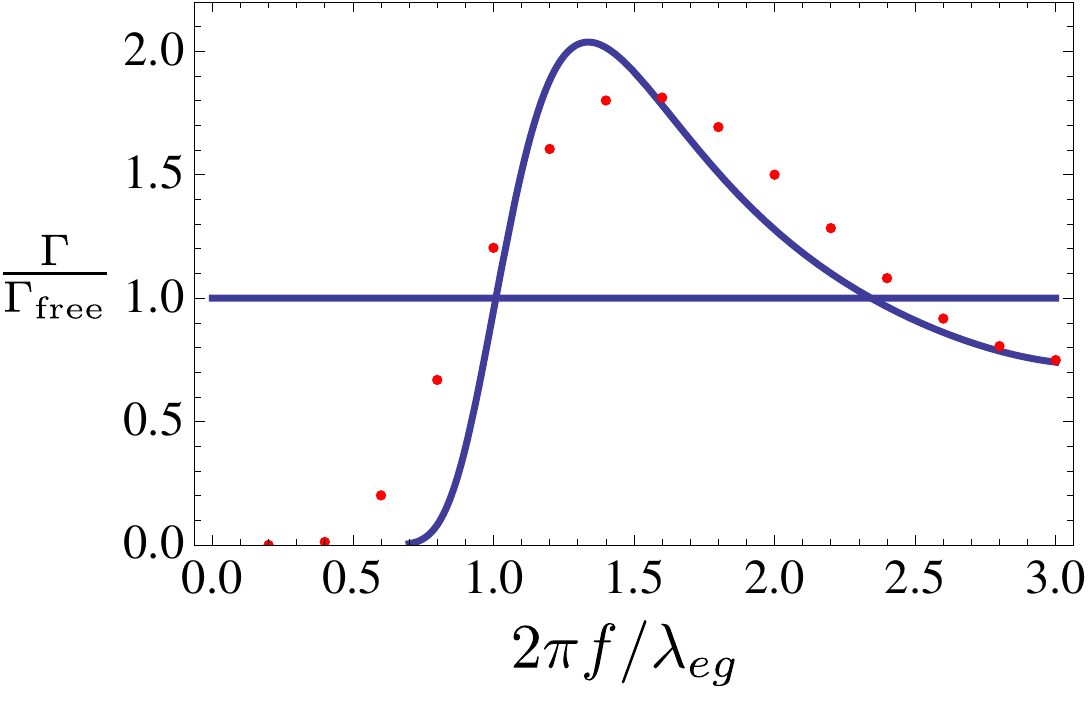}
		 \caption{Scaled spontaneous decay rate ${\Gamma}/{\Gamma_{\text{free}}}$ as a function of the scaled focal length $2\pi{f}/{\lambda_{eg}}$ of
the parabola: 
Numerically exact results (dots),
semiclassical results (solid).
We also indicate the free-space value as a horizontal line.
}
		 \label{fig:Purcell}

\end{figure}

We can use our results of the preceding chapter to calculate the influence of the Purcell effect \cite{Purcell} on the spontaneous decay rate. In the pole or Weisskopf-Wigner \cite{Wigner} approximation the spontaneous decay rate of atom $a$  is given by 
$\Gamma=2 \text{Re}\left[A^{a,a}\right]\Bigr\vert_{s=-\iE \omega_{eg}}\;.$
Thus, for a parabolic cavity we obtain the result \cite{Alber2013}
\begin{align*} 
\frac{\Gamma}{\Gamma_{\text{free}}}=1+2\sum\limits _{M=1}^{\infty}3\cos\left[2M(u-\pi/2)\right]\\
\times\frac{2M\mathcal{S}(u)\coth\left[2M\mathcal{S}(u)\right]-1}{\sinh^{2}\left[2M\mathcal{S}(u)\right]}\Bigr\vert_{u=2\pi\frac{f}{\lambda_{eg}}}
\end{align*}
with $\mathcal{S}(u):=\int\limits _{0}^{u}\sin^{2}(y)/y$ and with $f$ denoting the focal length of the parabola. 
Each summand in the expression for ${\Gamma}/{\Gamma_{\text{free}}}$ is connected with a possible photon path inside the cavity, i.e. with
bouncing back and forth of a photon (counted by $M$) between the atom and the vertex of the parabola. 
These contributions to the decay rate are connected with diffraction effects which are suppressed for large values of $f\gg\lambda_{eg}$. Similar but
 more complicated results are obtained for an ellipsoidal cavity.
The main difference between both cases is that for an ellipsoidal cavity we obtain a double sum which can be traced back to the existence of two quantization conditions.
Of course, the results for a parabolic cavity can be obtained from the more general results for an ellipsoidal cavity in the limit in which the distance between the two focal points tends to infinity.
In figure \ref{fig:Purcell} the semiclassical analytical results are compared with
exact numerical results.
This numerical comparison demonstrates the quality of the semiclassical results even in cases in which the wave length of the spontaneously emitted photon becomes comparable to the characteristic length scales inside the cavity. Thus,
the previously discussed semiclassical approximation also enables us to investigate diffraction effects which become relevant for $f \lesssim \lambda_{eg}$.
\begin{figure}
	\centering
		\includegraphics[width=5.5cm]{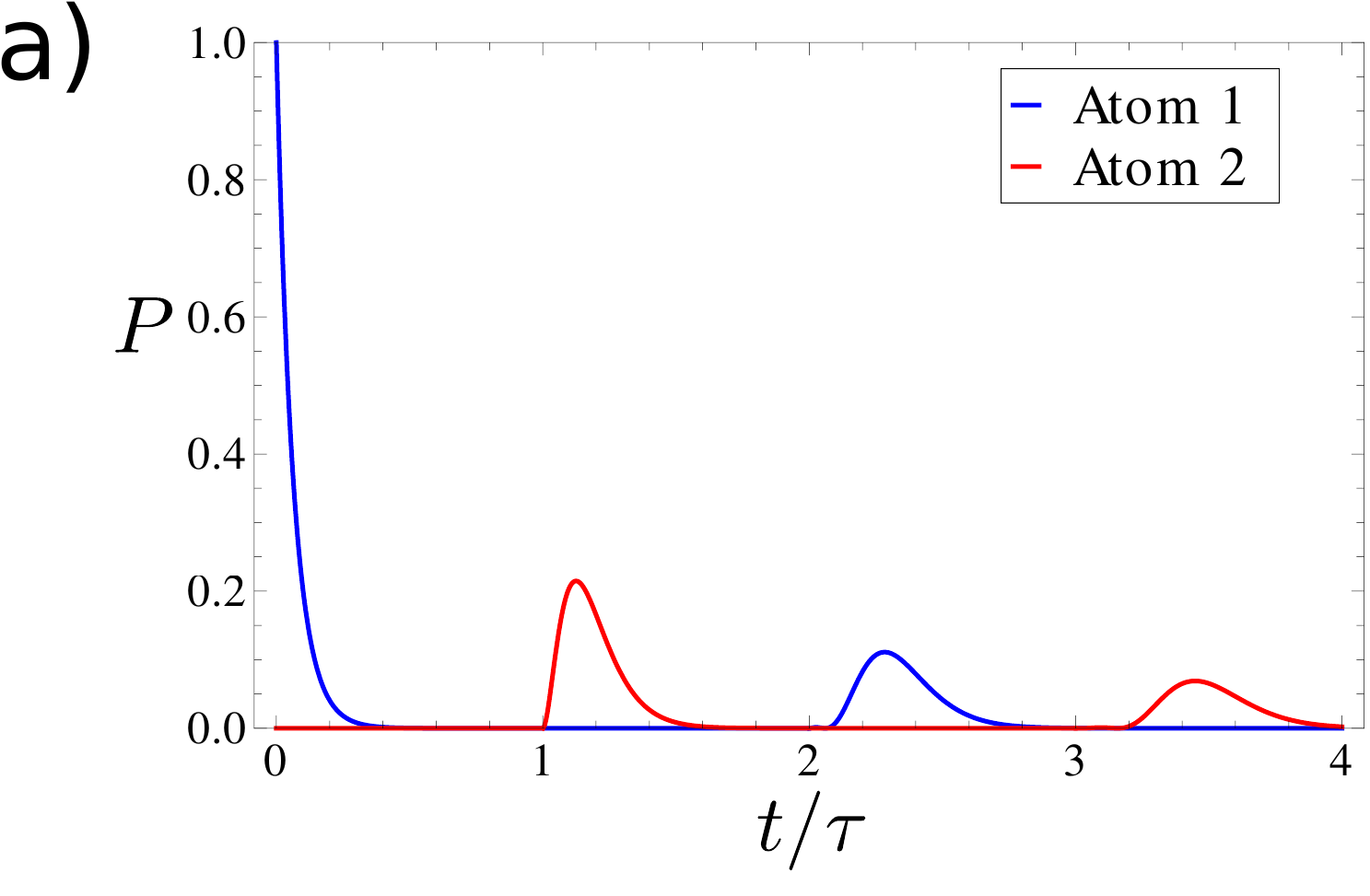}\\
		\includegraphics[width=5.5cm]{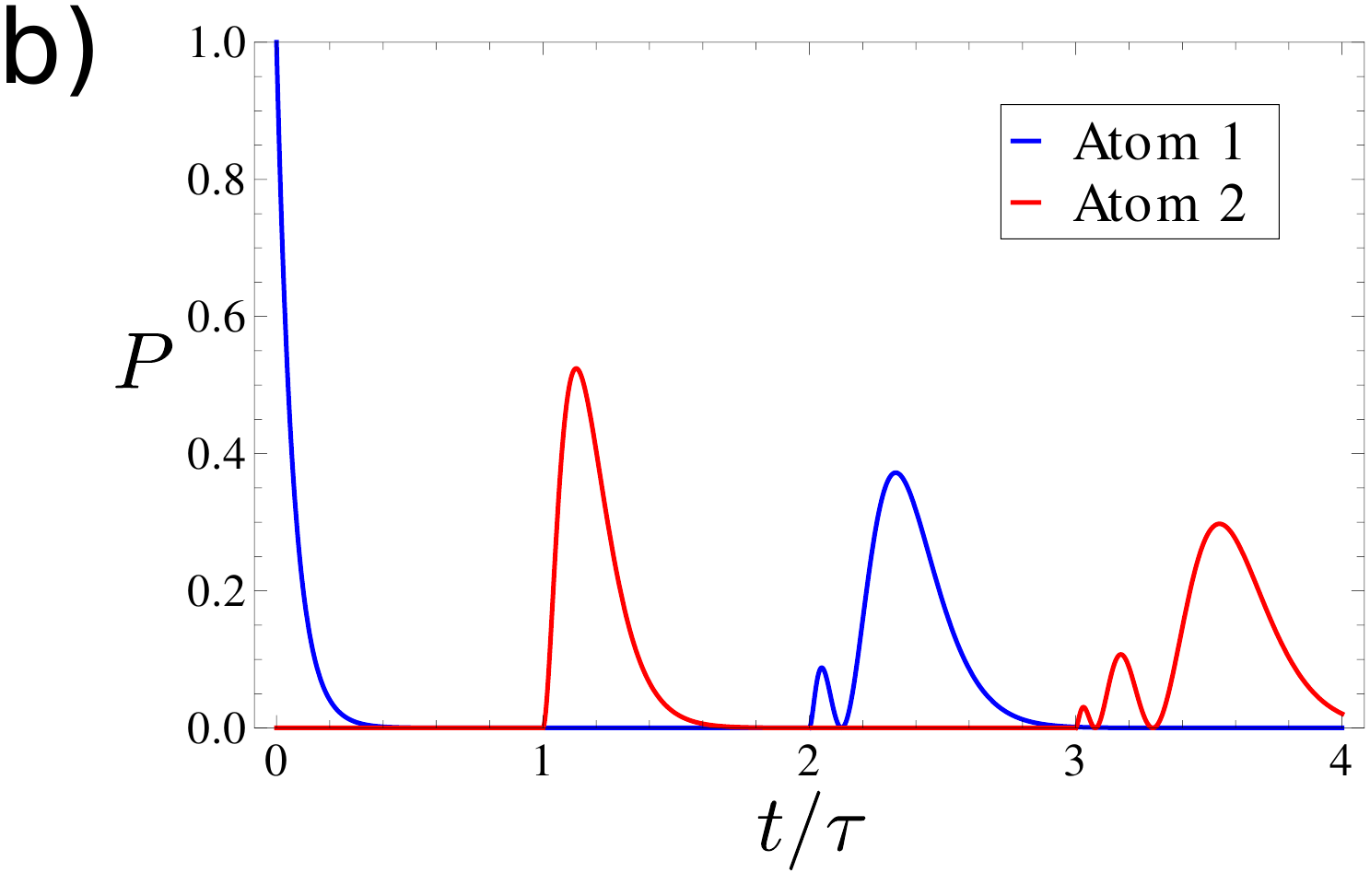}
		 \caption{Atomic excitation probabilities for $\Gamma_{\text{free}}\cdot\tau=16$ and ${f}/{d}={1}/{2}$ (a) and ${f}/{d}={9}/{2}$ (b). The excitation probability of atom
		$1$ ($2$) is represented by the blue (red) curve.}
		 \label{fig:Atomic_excitation_probabilities}

\end{figure}

\section{Time evolution of atomic excitation probabilities}
We can now investigate the time evolution of atomic excitations. In this chapter we are going to present results for a prolate  ellipsoidal cavity.
For this purpose we need some additional quantities for describing the cavity, namely the distance $d$ between the focal points and the shortest
distance $f$ between one of the vertices and its closest focal point. In the following we concentrate on
the regime of short wavelengths, i.e. $d, \,f\gg \lambda_{eg}$, in order to obtain a simple picture of the dynamics. If we start with an initial excitation of atom 1 we observe
a bouncing back and forth of the atomic excitation between the two atoms inside the cavity by exchange of a spontaneously emitted photon. The time $\tau$ the excitation needs to travel
from one atom to another is given by $\tau={(d+2 f)}/{c}$. The corresponding behavior of the atomic excitation probabilities is depicted in figures \ref{fig:Atomic_excitation_probabilities} a and b. 
We observe that the probability of a successful transfer of the excitation to the second atom depends significantly on the shape of the cavity.
If $f\gg d$, i.e. the cavity is almost spherically symmetric, the excitation probabilities
associated with repeated returns of a photon to an atom are significantly larger than for cases with $d\gg f$. This can be observed by comparing  the excitation probabilities depicted in figures \ref{fig:Atomic_excitation_probabilities} a) and
\ref{fig:Atomic_excitation_probabilities} b). 
This phenomenon may be traced back to the fact that
both atomic dipole matrix elements are oriented along the z-axis. In the case of  
$d\gg f$  the radiation field of the photon emitted spontaneously by the first atom
has to propagate
almost along the z-axis in order to approach atom $2$.
Therefore, its electrical field is almost perpendicular to the z-axis and its coupling to the dipole of atom 2 is small. As a consequence the excitation
probability is also small. In the opposite limit of an almost spherically symmetric cavity
such a confinement of the photonic wave packet along the symmetry axis due to the shape of the cavity 
does not occur and the excitation probabilities are larger.
\begin{figure*}[ht]
		\centering
			 \includegraphics[width=4cm]{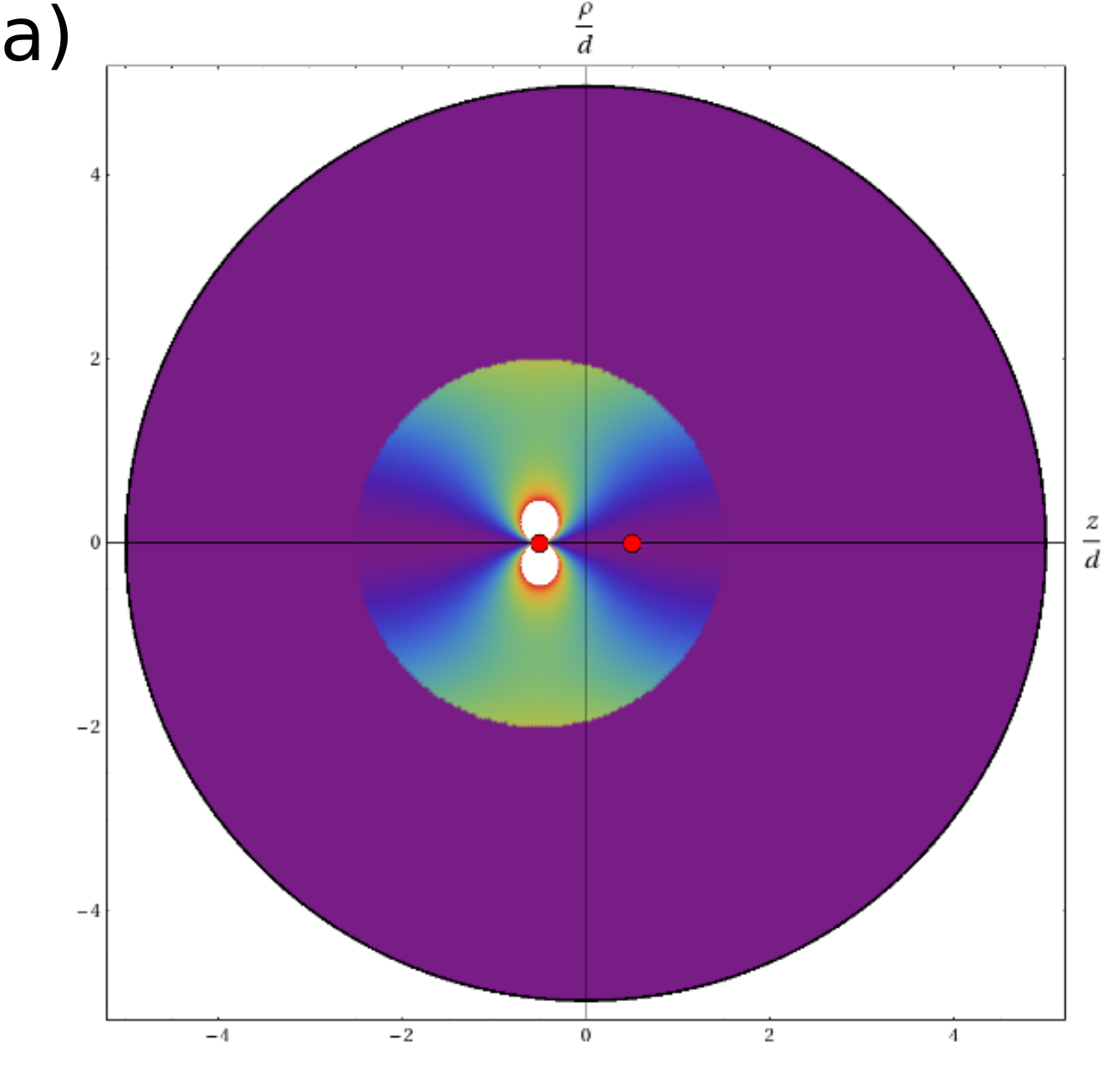}
			  \includegraphics[width=4cm]{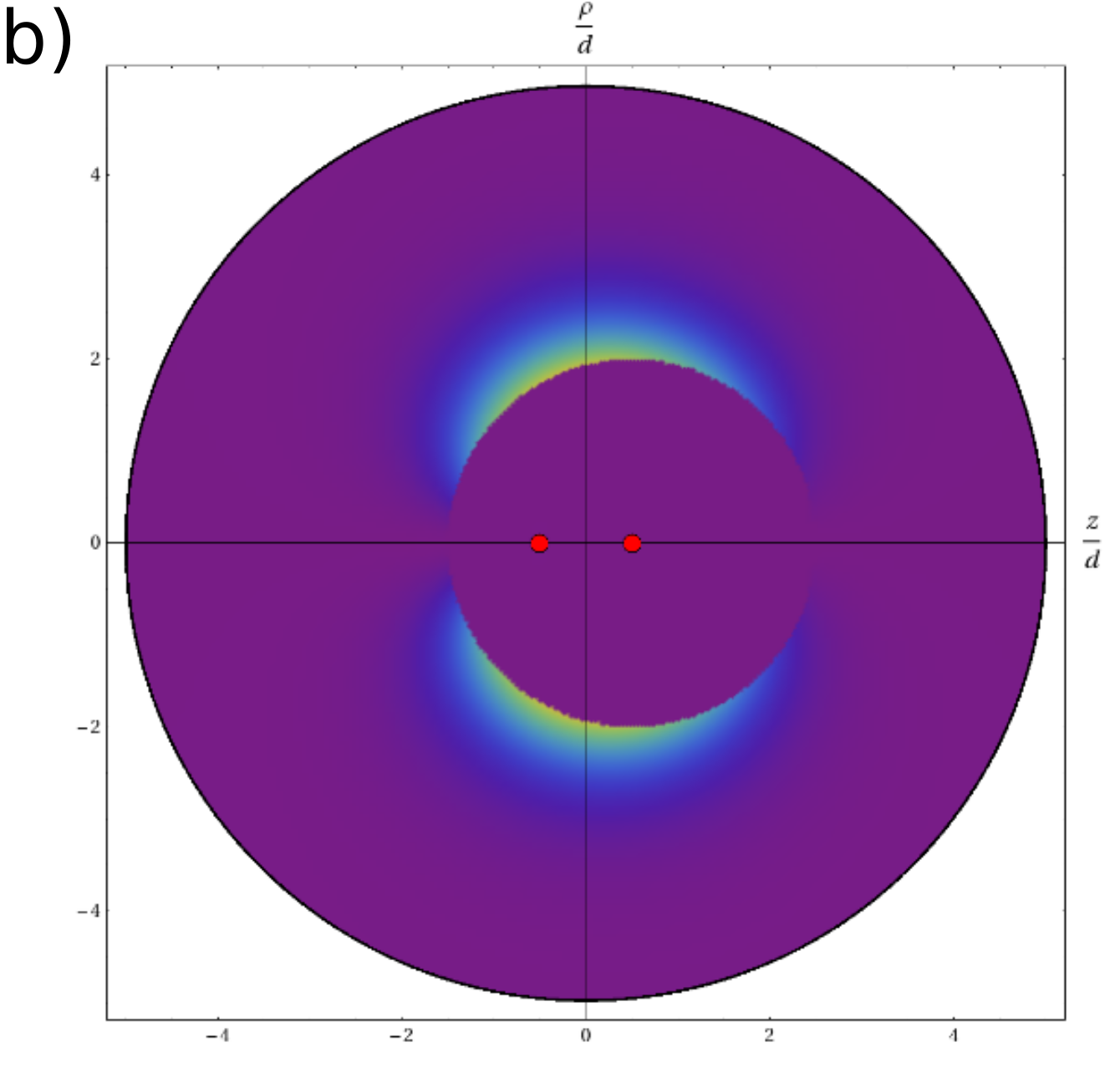}
			  \includegraphics[width=4cm]{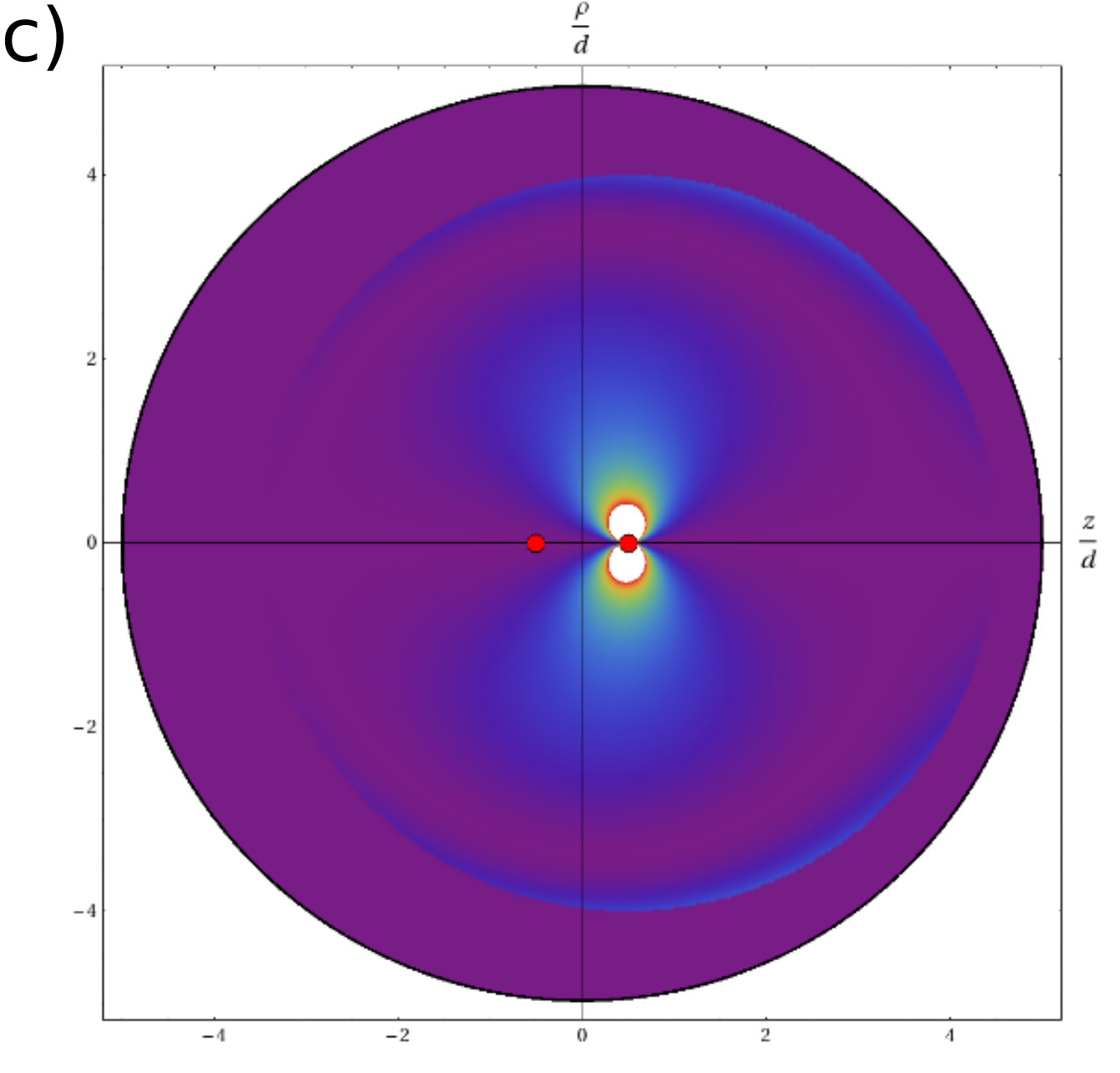}
			  \includegraphics[width=4cm]{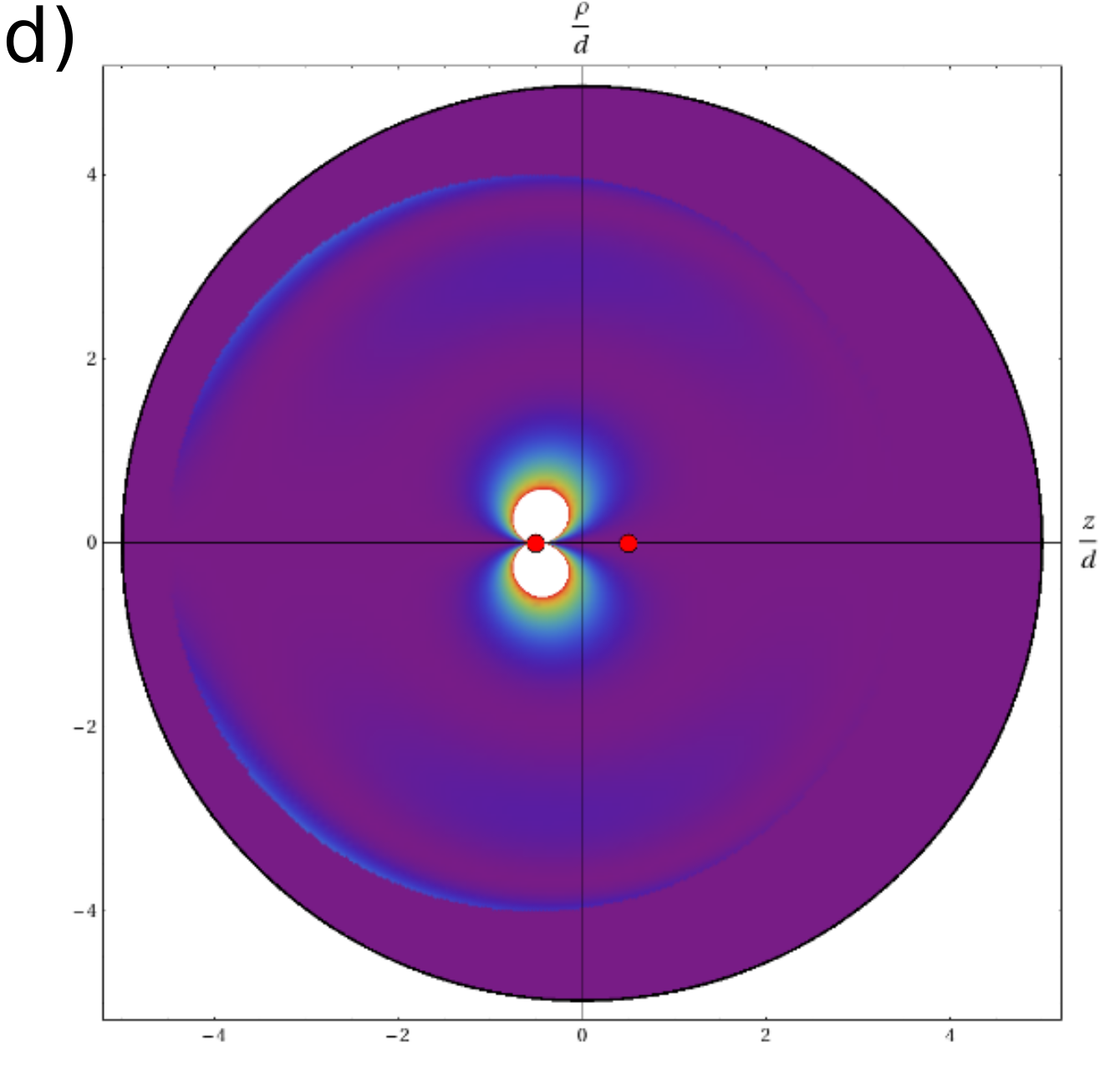}\\
			   \includegraphics[width=3cm]{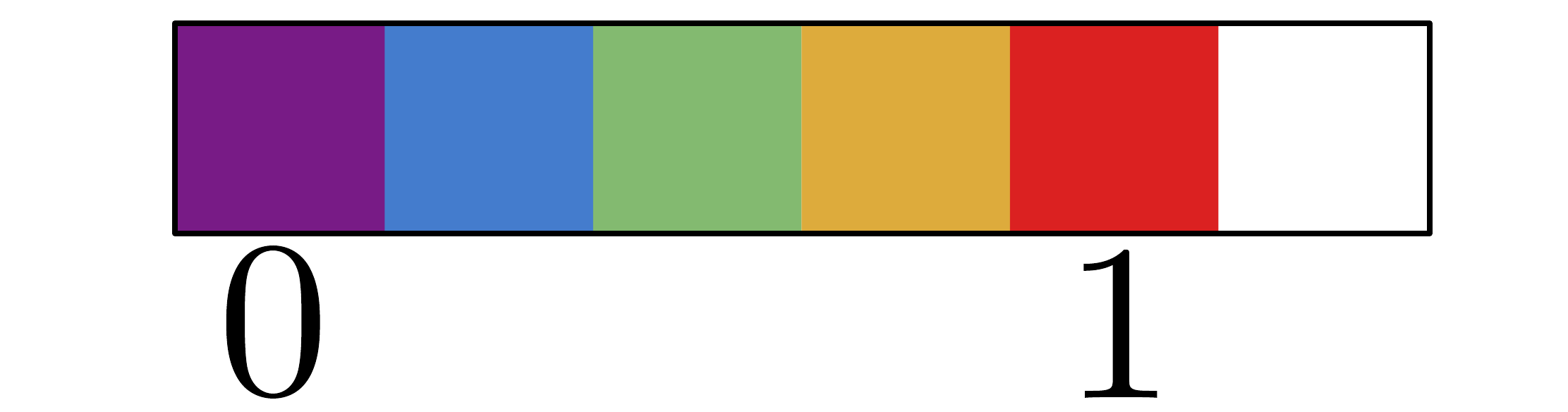}
			 \caption{Time evolution of the (normally ordered) energy density (in units of $3\pi\hbar\omega_{eg}\Gamma_{\text{free}}/(80d^{2}c)$) of a spontaneously emitted photon in a prolate ellipsoidal cavity: The parameters are $\Gamma_{\text{free}}\tau = 16$ and ${f}/{d}= 9/2$; 
			$t=0.2\tau$ (a),$t=0.8\tau$ (b), $t=1.4\tau$ (c), and $t=2.4\tau$ (d). The red dots indicate the focal points.}
			 \label{fig:Energydensity}

	\end{figure*}
\section{Time evolution of the energy density of the electromagnetic radiation field}
We can also calculate the energy density of the electromagnetic radiation field in the case $f,d\gg\lambda_{eg}$ by using multidimensional semiclassical methods \cite{Maslov1981}. Let us investigate the dynamics of a photonic wave packet which carries the excitation from one atom to the other inside a prolate ellipsoidal cavity. Similar to the bouncing back and forth in the time evolution of the atomic excitation probabilities we can observe a bouncing back
and forth  of the photonic wave packet in the cavity. The dynamics of repeated absorption and spontaneous emission events also leads to a change of the shape of the photonic wave packet.
This process  is illustrated in figures \ref{fig:Energydensity} a-d for the electromagnetic energy density of a single photon. The corresponding atomic excitation probabilities are depicted in figure \ref{fig:Atomic_excitation_probabilities} b.
Figure \ref{fig:Energydensity} a illustrates the energy density produced by the spontaneous emission of a single photon. The cavity guides the one-photon wave packet to the second focus as illustrated in
figure \ref{fig:Energydensity} b. Figure \ref{fig:Energydensity} c shows the wave packet directly after the interaction with the second atom. Figure  \ref{fig:Energydensity} d illustrates the wave packet after 
its interaction with the first atom.
	
\section{Conclusion}
We have investigated the dynamics of spontaneous photon emission of two-level systems in parabolic and prolate ellipsoidal cavities assuming that the atoms are situated in the focal points of these cavities.
Our theoretical approach is based on treating the multimode aspects of this problem with the help of one-dimensional semiclassical methods by exploiting the separability of the relevant Helmholtz equation. This way we are able to improve straight forward multidimensional semiclassical approaches considerably. With the help of semiclassical
photon path representations of relevant probability amplitudes we have explored the intricate
dynamical interplay between spontaneous photon emission and absorption processes on the one hand and
reflections of the photon wave packet at the boundary of the cavity on the other hand.


\begin{thebibliography}{15}%
\makeatletter
\providecommand \@ifxundefined [1]{%
 \@ifx{#1\undefined}
}%
\providecommand \@ifnum [1]{%
 \ifnum #1\expandafter \@firstoftwo
 \else \expandafter \@secondoftwo
 \fi
}%
\providecommand \@ifx [1]{%
 \ifx #1\expandafter \@firstoftwo
 \else \expandafter \@secondoftwo
 \fi
}%
\providecommand \natexlab [1]{#1}%
\providecommand \enquote  [1]{``#1''}%
\providecommand \bibnamefont  [1]{#1}%
\providecommand \bibfnamefont [1]{#1}%
\providecommand \citenamefont [1]{#1}%
\providecommand \href@noop [0]{\@secondoftwo}%
\providecommand \href [0]{\begingroup \@sanitize@url \@href}%
\providecommand \@href[1]{\@@startlink{#1}\@@href}%
\providecommand \@@href[1]{\endgroup#1\@@endlink}%
\providecommand \@sanitize@url [0]{\catcode `\\12\catcode `\$12\catcode
  `\&12\catcode `\#12\catcode `\^12\catcode `\_12\catcode `\%12\relax}%
\providecommand \@@startlink[1]{}%
\providecommand \@@endlink[0]{}%
\providecommand \url  [0]{\begingroup\@sanitize@url \@url }%
\providecommand \@url [1]{\endgroup\@href {#1}{\urlprefix }}%
\providecommand \urlprefix  [0]{URL }%
\providecommand \Eprint [0]{\href }%
\providecommand \doibase [0]{http://dx.doi.org/}%
\providecommand \selectlanguage [0]{\@gobble}%
\providecommand \bibinfo  [0]{\@secondoftwo}%
\providecommand \bibfield  [0]{\@secondoftwo}%
\providecommand \translation [1]{[#1]}%
\providecommand \BibitemOpen [0]{}%
\providecommand \bibitemStop [0]{}%
\providecommand \bibitemNoStop [0]{.\EOS\space}%
\providecommand \EOS [0]{\spacefactor3000\relax}%
\providecommand \BibitemShut  [1]{\csname bibitem#1\endcsname}%
\let\auto@bib@innerbib\@empty
\bibitem [{\citenamefont {Berman}(1994)}]{Berman1994}%
  \BibitemOpen
  \bibinfo {editor} {\bibfnamefont {P.}~\bibnamefont {Berman}},\ ed.,\
  \href@noop {} {\emph {\bibinfo {title} {Cavity Quantum Electrodynamics}}}\
  (\bibinfo  {publisher} {Academic Press},\ \bibinfo {address} {San Diego},\
  \bibinfo {year} {1994})\BibitemShut {NoStop}%
\bibitem [{\citenamefont {Walther}\ \emph {et~al.}(2006)\citenamefont {Walther}
  \emph {et~al.}}]{Walther2006}%
  \BibitemOpen
  \bibfield  {author} {\bibinfo {author} {\bibfnamefont {H.}~\bibnamefont
  {Walther}, \bibfnamefont {H.}} \emph {et~al.},\ }\href@noop {} {\bibfield
  {journal} {\bibinfo  {journal} {Prog. Phys.}\ }\textbf {\bibinfo {volume}
  {69}},\ \bibinfo {pages} {1325} (\bibinfo {year} {2006})}\BibitemShut
  {NoStop}%
\bibitem [{\citenamefont {Haroche}\ \emph {et~al.}(2006)\citenamefont {Haroche}
  \emph {et~al.}}]{Haroche2006}%
  \BibitemOpen
  \bibfield  {author} {\bibinfo {author} {\bibfnamefont {S.}~\bibnamefont
  {Haroche}} \emph {et~al.},\ }\href@noop {} {\emph {\bibinfo {title}
  {Exploring the Quantum: Atoms, Cavities and Photons}}}\ (\bibinfo
  {publisher} {Oxford University Press},\ \bibinfo {address} {Oxford},\
  \bibinfo {year} {2006})\BibitemShut {NoStop}%
\bibitem [{\citenamefont {Goy}\ \emph {et~al.}(1983)\citenamefont {Goy} \emph
  {et~al.}}]{Goy1983}%
  \BibitemOpen
  \bibfield  {author} {\bibinfo {author} {\bibfnamefont {P.}~\bibnamefont
  {Goy}} \emph {et~al.},\ }\href@noop {} {\bibfield  {journal} {\bibinfo
  {journal} {Phys. Rev. Lett.}\ }\textbf {\bibinfo {volume} {50}},\ \bibinfo
  {pages} {1903} (\bibinfo {year} {1983})}\BibitemShut {NoStop}%
\bibitem [{\citenamefont {Meschede}\ \emph {et~al.}(1985)\citenamefont
  {Meschede} \emph {et~al.}}]{Meschede1985}%
  \BibitemOpen
  \bibfield  {author} {\bibinfo {author} {\bibfnamefont {D.}~\bibnamefont
  {Meschede}} \emph {et~al.},\ }\href@noop {} {\bibfield  {journal} {\bibinfo
  {journal} {Phys. Rev. Lett.}\ }\textbf {\bibinfo {volume} {54}},\ \bibinfo
  {pages} {551} (\bibinfo {year} {1985})}\BibitemShut {NoStop}%
\bibitem [{\citenamefont {McKeever}\ \emph {et~al.}(2003)\citenamefont
  {McKeever} \emph {et~al.}}]{McKeever2003}%
  \BibitemOpen
  \bibfield  {author} {\bibinfo {author} {\bibfnamefont {J.}~\bibnamefont
  {McKeever}} \emph {et~al.},\ }\href@noop {} {\bibfield  {journal} {\bibinfo
  {journal} {Nature}\ }\textbf {\bibinfo {volume} {425}},\ \bibinfo {pages}
  {268} (\bibinfo {year} {2003})}\BibitemShut {NoStop}%
\bibitem [{\citenamefont {Schleich}(2001)}]{Schleich}%
  \BibitemOpen
  \bibfield  {author} {\bibinfo {author} {\bibfnamefont {W.}~\bibnamefont
  {Schleich}},\ }\href@noop {} {\emph {\bibinfo {title} {Quantum Optics in
  Phase Space}}}\ (\bibinfo  {publisher} {Wiley-VCH},\ \bibinfo {address}
  {Berlin},\ \bibinfo {year} {2001})\BibitemShut {NoStop}%
\bibitem [{\citenamefont {Maiwald}\ \emph {et~al.}(2009)\citenamefont {Maiwald}
  \emph {et~al.}}]{Maiwald2009}%
  \BibitemOpen
  \bibfield  {author} {\bibinfo {author} {\bibfnamefont {R.}~\bibnamefont
  {Maiwald}} \emph {et~al.},\ }\href@noop {} {\bibfield  {journal} {\bibinfo
  {journal} {Nat. Phys.}\ }\textbf {\bibinfo {volume} {5}},\ \bibinfo {pages}
  {551 } (\bibinfo {year} {2009})}\BibitemShut {NoStop}%
\bibitem [{\citenamefont {Maiwald}\ \emph {et~al.}(2012)\citenamefont {Maiwald}
  \emph {et~al.}}]{Maiwald2012}%
  \BibitemOpen
  \bibfield  {author} {\bibinfo {author} {\bibfnamefont {R.}~\bibnamefont
  {Maiwald}} \emph {et~al.},\ }\href@noop {} {\bibfield  {journal} {\bibinfo
  {journal} {Phys. Rev. A}\ }\textbf {\bibinfo {volume} {86}},\ \bibinfo
  {pages} {043431} (\bibinfo {year} {2012})}\BibitemShut {NoStop}%
\bibitem [{\citenamefont {Moon}\ and\ \citenamefont
  {Eberle-Spencer}(1961)}]{Moon1961}%
  \BibitemOpen
  \bibfield  {author} {\bibinfo {author} {\bibfnamefont {P.}~\bibnamefont
  {Moon}}\ and\ \bibinfo {author} {\bibfnamefont {D.}~\bibnamefont
  {Eberle-Spencer}},\ }\href@noop {} {\emph {\bibinfo {title} {Field Theory
  Handbook}}}\ (\bibinfo  {publisher} {Springer},\ \bibinfo {address}
  {Berlin},\ \bibinfo {year} {1961})\BibitemShut {NoStop}%
\bibitem [{\citenamefont {Milonni}\ and\ \citenamefont
  {Knight}(1974)}]{Milonni1974}%
  \BibitemOpen
  \bibfield  {author} {\bibinfo {author} {\bibfnamefont {P.~W.}\ \bibnamefont
  {Milonni}}\ and\ \bibinfo {author} {\bibfnamefont {P.~L.}\ \bibnamefont
  {Knight}},\ }\href@noop {} {\bibfield  {journal} {\bibinfo  {journal} {Phy.
  Rev. A}\ }\textbf {\bibinfo {volume} {10}},\ \bibinfo {pages} {1096}
  (\bibinfo {year} {1974})}\BibitemShut {NoStop}%
\bibitem [{\citenamefont {Alber}\ \emph {et~al.}(2013)\citenamefont {Alber}
  \emph {et~al.}}]{Alber2013}%
  \BibitemOpen
  \bibfield  {author} {\bibinfo {author} {\bibfnamefont {G.}~\bibnamefont
  {Alber}} \emph {et~al.},\ }\href@noop {} {\bibfield  {journal} {\bibinfo
  {journal} {Phy. Rev. A}\ }\textbf {\bibinfo {volume} {88}},\ \bibinfo {pages}
  {023825} (\bibinfo {year} {2013})}\BibitemShut {NoStop}%
\bibitem [{\citenamefont {Purcell}(1946)}]{Purcell}%
  \BibitemOpen
  \bibfield  {author} {\bibinfo {author} {\bibfnamefont {E.}~\bibnamefont
  {Purcell}},\ }\href@noop {} {\bibfield  {journal} {\bibinfo  {journal} {Phys.
  Rev.}\ }\textbf {\bibinfo {volume} {69}},\ \bibinfo {pages} {681} (\bibinfo
  {year} {1946})}\BibitemShut {NoStop}%
\bibitem [{\citenamefont {Weisskopf}\ and\ \citenamefont
  {Wigner}(1930)}]{Wigner}%
  \BibitemOpen
  \bibfield  {author} {\bibinfo {author} {\bibfnamefont {V.}~\bibnamefont
  {Weisskopf}}\ and\ \bibinfo {author} {\bibfnamefont {E.}~\bibnamefont
  {Wigner}},\ }\href@noop {} {\bibfield  {journal} {\bibinfo  {journal} {Z.
  Physik}\ }\textbf {\bibinfo {volume} {63}},\ \bibinfo {pages} {54} (\bibinfo
  {year} {1930})}\BibitemShut {NoStop}%
\bibitem [{\citenamefont {Maslov}\ and\ \citenamefont
  {Fedoriuk}(1981)}]{Maslov1981}%
  \BibitemOpen
  \bibfield  {author} {\bibinfo {author} {\bibfnamefont {V.~P.}\ \bibnamefont
  {Maslov}}\ and\ \bibinfo {author} {\bibfnamefont {M.~V.}\ \bibnamefont
  {Fedoriuk}},\ }\href@noop {} {\emph {\bibinfo {title} {Semi-Classical
  Approximation in Quantum Mechanics}}}\ (\bibinfo  {publisher} {D. Reidel},\
  \bibinfo {address} {Dordrecht},\ \bibinfo {year} {1981})\BibitemShut
  {NoStop}%
\end{thebibliography}
\end{document}